\documentclass[epj]{svjour}
\usepackage{graphics}
\begin{document}
\title{Anisotropic generalization of Vaidya-Tikekar superdense star}
\author{S. Thirukkanesh\inst{1}\thanks{Email: thirukkanesh@esn.ac.lk} \and Ranjan Sharma\inst{2}\thanks{E-mail: rsharma@associates.iucaa.in} \and Sunil D. Maharaj\inst{3}\thanks{E-mail: maharaj@ukzn.ac.za}
}                     
\offprints{}          

\institute{Department of Mathematics, Eastern University, Chenkalady, Sri Lanka \\ 
\and  Department of Physics, Cooch Behar Panchanan Barma University, Cooch Behar 736101, India\\
 \and Astrophysics and Cosmology Research Unit, School of Mathematics,
Statistics and Computer Science, University of KwaZulu-Natal,
Private Bag X54001, Durban 4000, South Africa.\\}
\date{Received: date / Revised version: date}
%
\abstract{We study superdense relativistic stars with anisotropic matter distributions with spheroidal spatial hypersurfaces.
We propose a methodology to make an anisotropic generalization of the Vaidya-Tikekar superdense star model. 
The anisotropic Einstein field equations can be solved in terms of hypergeometric functions for our choice of gravitational potential
and anisotropy. Particular parameter choices allow us to generate models of anisotropic stars in terms of elementary
functions. Also, isotropic stars can be generated in the limit of vanishing anisotropy. In particular, we obtain the well known
superdense models of Tikekar which are isotropic and have specific spheroidal geometries. 
The impact of anisotropy on the gross physical behaviour of a compact star is studied.
} 
\maketitle

\section{\label{sec1} Introduction}
In astrophysics, theoretical modelling of neutron stars began much before its actual discovery\cite{Shapiro}. The standard approach to analyze the physical properties of a neutron star is to solve the Tolman-Oppenheimer-Volkoff equations for a given equation of state (EOS). On the other hand, finding exact solutions to Einstein field equations for stellar fluid distributions and interpreting them are useful avenues in understanding the properties of self-gravitating objects. However, on top of the problem of making the system of equations tractable, physical acceptability poses a huge challenge in this direction and so far only a limited number of exact solutions have been found which are well behaved, regular and can describe realistic stars like neutron stars (see Ref.~\cite{Delgaty98}). Given this background, different ad hoc approaches are adopted to make the system of equations tractable, and their viability explored. One such approach that has got huge recognition amongst researchers is the ansatz of Vaidya and Tikekar\cite{Vaidya83}. In this approach, we assume  a particular spatial geometry for the $t =$ constant hypersurface of a static spherically symmetric star which turns out to be spheroidal rather than usual spherical geometry. The resultant solutions for specific values of the curvature parameter associated with the spheroidal geometry $K$ are shown to be useful for the description of compact stellar objects like neutron stars\cite{Tikekar90}. Later on, solutions for different values of the curvature parameter have been obtained which include polynomial solutions\cite{Maharaj96} and a general solution\cite{Mukherjee97}. Komathiraj and Maharaj\cite{Komathiraj} extended the polynomial solutions to include the electromagnetic field. Since the Vaidya-Tikekar ansatz was shown to be relevant for developing physically realizable stellar models, the technique has attracted huge research interests and subsequently, many papers have been published making use of the ansatz.
There have been many models of physically acceptable compact spheres in general relativity which relate  to specific spheroidal parameters. Examples include the  astrophysical models of Paul {\em et al}\cite{newref1} and Chattopadhyay {\em et al}\cite{newref2}. Note that spheroidal spacetimes have been applied to mixtures involving quark-diquark structures, models of strange stars, compact objects with equations of state, relativistic core-envelope models and general relativistic radiating stars undergoing  dissipation. Bhar\cite{newref3} found physically acceptable models with spheroidal hypersurfaces with quintessence and containing a conformal Killing vector, and Hansraj\cite{newref4} demonstrated that  the five-dimensional generalization of spheroidal spacetimes is applicable to dense objects in Einstein-Gauss-Bonnet gravity.

Keeping in mind the huge success of the Vaidya-Tikekar ansatz, we plan to make a further generalization of the class of solutions developed earlier. Note that a natural extension of the Vaidya-Tikekar stellar model can be done by incorporating the electromagnetic field\cite{Sharma01}. Another important parameter that has found its relevance in the studies of relativistic compact stars is the presence of anisotropy (radial pressure different from tangential pressure). It is well established that physical properties of compact stars do get  significantly influenced if there exist unequal stresses locally. The microscopic origin and the effects of local pressure anisotropy in astrophysical systems have been extensively analyzed by Herrera and Santos\cite{Herrera1}. Recently, some of us have developed an anisotropic generalization of the Finch and Skea\cite{Finch89} stellar model which has been found to be very useful in the context of realistic compact stars\cite{Thiru17}. In another work, we have prescribed an algorithm to generate a new class of exact solutions by relaxing the pressure isotropy condition and showed that the new solutions are anisotropic generalizations of a large class of isotropic stellar models\cite{Thiru18}. Anisotropy may develop due to a large variety of physical phenomena we expect to find in systems like the high-density regime of compact stars. Details of microscopic origin of anisotropy are available in Ref.~\cite{Thiru17,Thiru18} and references therein. In this paper, we aim to provide a mechanism to generate a new class of solutions which may be treated as anisotropic generalizations of the Vaidya-Tikekar model.

\section{\label{sec2}The field equations}
We couch the interior spacetime of a static and spherically symmetric relativistic dense star (in coordinates $(x^{a}) =
(t,r,\theta,\phi)$) in the form
\begin{equation}
\label{1} ds^{2} = -e^{2\nu(r)} dt^{2} + e^{2\lambda(r)} dr^{2} +
 r^{2}(d\theta^{2} + \sin^{2}{\theta} d\phi^{2}).
\end{equation}
We assume the matter distribution of the dense star to be anisotropic in nature and accordingly the energy momentum tensor is assumed to be of the form
\begin{equation}
\label{2} T^i_{j}=\mbox{diag}(-\rho, p_r, p_t, p_t).
\end{equation}
The energy density $\rho$, the radial pressure $p_r$ and  the tangential pressure $p_t$ are measured relative to the comoving
fluid velocity $u^i = e^{-\nu}\delta^i_0.$  For the line element (\ref{1}) and matter distribution (\ref{2}), the Einstein field
equations (in system of units having $8\pi G = 1 = c$) can be expressed as
\begin{eqnarray}
\label{3} \rho &=& \frac{1}{r^{2}} \left[ r(1-e^{-2\lambda})
\right]',\\
\label{4} p_r &=& - \frac{1}{r^{2}} \left( 1-e^{-2\lambda} \right)
+
\frac{2\nu'}{r}e^{-2\lambda} ,\\
\label{5}p_t &=& e^{-2\lambda}\left( \nu'' + \nu'^{2} +
\frac{\nu'}{r}- \nu'\lambda' - \frac{\lambda'}{r} \right) ,
\end{eqnarray}
where a prime ($'$) denotes differentiation with respect to $r$. The system of equations (\ref{3})-(\ref{5}), determines the gravitational behaviour of the anisotropic imperfect fluid sphere. The mass contained within a radius $r$ of the sphere is defined as
\begin{equation}
\label{6} m(r)= \frac{1}{2}\int_0^r\omega^2 \rho(\omega)d\omega .
\end{equation}
A different but equivalent form of the field equations can be found if we introduce the Durgapal and Bannerji\cite{Durgapal83} transformation
\begin{equation}
\label{7} x = Cr^2,~~ Z(x)  = e^{-2\lambda(r)} ~\mbox{and}~
A^{2}y^{2}(x) = e^{2\nu(r)},
\end{equation}
where $A$ and $C> 0$ are  arbitrary constants.  Subsequently, the line element (\ref{1}) takes the form
\begin{equation}
\label{8} ds^2 = -A^2 y^2 dt^2 + \frac{1}{4CxZ}dx^2 + \frac{x}{C}(d\theta^2+\sin^2\theta d\phi^2).
\end{equation}
Under the transformation (\ref{7}), the equivalent system of equations  (\ref{3})-(\ref{5}) can be expressed as
\begin{eqnarray}
\label{9}  \frac{\rho}{C}&=& \frac{1-Z}{x} - 2\dot{Z}   , \\
\label{10} \frac{p_r}{C}&=& 4Z\frac{\dot{y}}{y} + \frac{Z-1}{x}
, \\
\label{11} p_t &=& p_r +\Delta , \\
\label{12}0 & =& 4x^2 Z\ddot{y}+ 2x^2 \dot{Z}\dot{y} + \left(x
\dot{Z} -Z+1-\frac{\Delta x}{C}\right)y,
\end{eqnarray}
where $\Delta = p_t-p_r$ is the measure of anisotropy and a dot ($.$) denotes differentiation with respect to the variable $x$. The mass
function (\ref{6}) in terms of the new variables in (\ref{7}) now takes the form
\begin{equation}
\label{13}m(x)=\frac{1}{4C^{3/2}} \int_0^x\sqrt{w}\rho(w)dw.
\end{equation}

\section{\label{sec3} Technique to generate new solutions}
Note that the system  (\ref{3})-(\ref{5}) comprises three independent equations in five unknowns $Z,~y,~\rho,~p_r,~p_t$. Naturally  the equivalent system (\ref{9})-(\ref{12}) can be solved if two of these unknowns are assumed a priori. We seek to solve the system by making explicit choices for the
gravitational potential $Z$ and the anisotropic parameter $\Delta$. Accordingly, we make the following assumptions
\begin{eqnarray}
\label{14} Z &=& \frac{1-x}{1-Kx}, \\
\label{15} \Delta &=& \frac{a K C  x}{(1 -Kx)^2},
\end{eqnarray}
where $a$ and $K$ are real constants. The choice (\ref{14}) is non-singular at the origin and was used  earlier by Vaidya and Tikekar \cite{Vaidya83} to study neutron stars having isotropic fluid distributions. Geometric interpretations of the choice and its physical implications may be found in reference \cite{Vaidya83} and subsequent works \cite{Maharaj96,Mukherjee97}. 
Effectively the parameter  $K$ allows for a wider
range of spatial geometries.
As far as the second choice is concerned, it is a reasonable assumption in the sense that $\Delta$ vanishes at the center (i.e., $p_r=p_t$ at the origin) which is consistent with the physical requirement for a realistic stellar model. Substitution of (\ref{14})  and (\ref{15}) in equation (\ref{12}) yields
\begin{equation}
\label{16} 4(1-x)(1-Kx) \ddot{y}-2(K-1) \dot{y}+ K(K-1- a)y = 0.
\end{equation}
To integrate (\ref{16}), we now introduce the following transformation
\begin{equation}
\label{17}  X = \frac{1}{1-K}(1-Kx),~~~ Y(X) = y(x),
\end{equation}
so that equation (\ref{16}) takes the form
\begin{equation}
\label{18} X(X-1) \frac{d^{2}Y}{dX^{2}}+\frac{1}{2}\frac{dY}{dX} -
\frac{(1-K+a)}{4}Y = 0,
\end{equation}
which is a Gaussian type hypergeometric equation. Note that in the context of anisotropic stellar modelling, special cases of this particular equation were earlier considered by Mak and Harko \cite{Mak03} and Harko and Mak \cite{Harko02} which can be obtained by setting $1-K+a = $constant and $1-K+a = 3$, respectively. 

The general solution of the equation is obtained as
\begin{equation}
\label{19} Y = C_1 F\left(\alpha, -(\alpha +1) ,-\frac{1}{2};X\right) + C_2X^{3/2}F\left(\frac{3}{2}+\alpha,
\frac{1}{2}-\alpha , \frac{5}{2};X\right),
\end{equation}
in terms of hypergeometric functions where $C_1,~C_2$ are constants
of integration and $\displaystyle \alpha = [-1 \pm
\sqrt{(2-K+a)}]/2$. In general (\ref{19}) can be written in the series
form as
\begin{equation}
\label{19a} Y = C_1 \left[1+ \sum_{j=1}^{\infty}  \frac{(\alpha)_j (-\alpha -1)_j}{( -\frac{1}{2})_j}\frac{X^j}{j!}\right] +
C_2X^{3/2}\left[1+ \sum_{j=1}^{\infty} \frac{(\frac{3}{2}+
\alpha)_j (\frac{1}{2} -\alpha )_j}{(
\frac{5}{2})_j}\frac{X^j}{j!}\right],
\end{equation}
where $(\alpha)_j=\alpha (\alpha +1)... (\alpha +j-1)$. It is
interesting to note that, for particular values of $K$ and $a$,
the series solution (\ref{19a}) can be expressed in terms of
elementary functions. This is possible, in general, as the
series terminates for specific values of the
model parameters $K$ and $a$. Using this feature, we present here two categories
of closed form solutions.

\section{\label{sec4} Exact Solutions}
Exact solutions  to (\ref{19a})  can be found in terms of elementary functions.
We present the two categories of solution possible.

\subsection{First category}
If we set $K-a=2 -(2n-1)^2$ (i.e.,  $\alpha =n-1$) and make use of the
properties of hypergeometric functions \cite{Polyania} then the solution (\ref{19})
can be written as
\begin{equation}
\label{19b} Y = C_1 F\left(n-1, -n, -\frac{1}{2};X\right) + C_2X^{3/2}
(1-X)^{1/2}F\left(2-n,n+1, \frac{5}{2};X\right),
\end{equation}
which yields
\begin{eqnarray}
\label{19c} && Y=  C_1  +  C_1 \sum_{j=1}^{n}  \frac{(-1)^{j-1}
2^{2j-1}n(n-1)}{j}\frac{(n+j-2)!X^j}{
(n-j)!(2j-2)!} + C_2X^{3/2} (1-X)^{1/2} \times \nonumber\\
&&\left[ 1+ \sum_{j=1}^{n-2} \frac{(-1)^j
2^{2j+2}3(j+2)(j+1)}{n(n-1)}  \times \frac{(n+j)!X^j}{(n-j-2)!(2j+4)!}
\right],
\end{eqnarray}
for  $n\geq 2$. 

\subsection{Second category}
If we set $K-a=2- 4n^2$ (i.e.,  $\alpha =(2n-1)/2$) and use  the properties of hypergeometric
functions \cite{Polyania} then the solution (\ref{19}) can be written as
\begin{equation}
\label{19d}  Y = C_1 (1-X)^{1/2} F\left(-n, n, -\frac{1}{2};X\right) + C_2X^{3/2} F\left(n+1, 1-n,
\frac{5}{2};X\right),
\end{equation}
which yields
\begin{eqnarray}
\label{19e} Y &=&  C_1 (1-X)^{1/2} \times\left[1+ \sum_{j=1}^{n}
\frac{(-1)^{j+1} 2^{2j-1}n}{j}\frac{(n+j-1)!X^j}{
(n-j)!(2j-2)!}\right]\nonumber\\
&&  + C_2X^{3/2}\left[1+\sum_{j=1}^{n-1} \frac{(-1)^j
2^{2j+2}3(j+2)(j+1)}{n}\times\frac{(n+j)!X^j}{(n-j-1)!(2j+4)!} \right],
\end{eqnarray}
for $n\geq 1$.

\section{\label{sec5} Particular models:}
Note that, making use of the series solutions given in (\ref{19c}) and (\ref{19e}), a large class of solutions can be obtained as anisotropic extensions to the Tikekar superdense stellar models. We demonstrate this by generating here some new solutions by specifying particular values to the model parameters:

\subsection{Case I: $n=2$ (i.e., $a-K=7, ~\alpha =1$)}
In this case, using (\ref{19c}), we obtain
\begin{equation}
\label{21} Y= C_1 \left[1+4X-8X^2\right]   + C_2 X^{\frac{3}{2}}
\sqrt{1-X}.
\end{equation}
In terms of the variable $x$, this solution eventually takes the form
\begin{equation}
\label{22} y = d_1 \left[1 + \frac{4}{(8-a)}[1+(7-a)x] -
\frac{8}{(8-a)^2} [1+(7-a)x]^2\right] + d_2
\left[\sqrt{1-x}[1+(7-a)x]^{3/2}\right],
\end{equation}
where we have introduced $d_1$ and $d_2$ as new arbitrary constants.

Consequently, the solution to the system (\ref{3})-(\ref{5}) can be written as
\begin{eqnarray}
e^{2\lambda} &=& \frac{1-(a-7)x}{1-x}, \\
e^{2\nu} &=& A^2y^2, \\
\frac{\rho}{C} &=& \frac{(8-a)[3+(7-a)x]}{[1+(7-a)x]^2}, \\
\frac{p_r}{C} &=& 4 \frac{(1-x)}{[1-(a-7)x]}\frac{\dot{y}}{y}-\frac{8-a}{[1+(7-a)x]}, \\
p_t &=& p_r+\Delta , \\
\Delta &=& \frac{a(a-7)Cx}{[1-(a-7)x]^2},
\end{eqnarray}
where $y$ is given in (\ref{22}).\\

Note that we can regain the Tikekar superdense stellar model for a particular value of $K$.
It is interesting to note that if we set $a=0, ~C=1/R^2$ and  $\displaystyle \tilde{x}=
\sqrt{1-x}=\sqrt{1-\frac{r^2}{R^2}}$, the solution (\ref{22}) reduces to
\begin{equation}
\label{23} y= d_3 \tilde{x} \left(1- \frac{7}{8}
\tilde{x}^2\right)^{3/2} +d_4 \left(1- \frac{7}{2}\tilde{x}^2
+\frac{49}{24}\tilde{x}^4\right),
\end{equation}
which is the Tikekar model \cite{Tikekar90} for a superdense neutron star for the spheroidal parameter
$K=-7$. The solution provided here is an anisotropic extension of the previous model.

\subsection{Case II: $n=1$ (i.e., $a-K=2, ~\alpha =\frac{1}{2}$)}
In this case, using (\ref{19e}), we obtain
\begin{equation}
\label{24} Y= C_1(1-X)^{1/2} (1+2X) + C_2 X^{\frac{3}{2}}.
\end{equation}
In terms of the variable $x$, the solution then takes the form
\begin{equation}
\label{25} y = e_1 \sqrt{(1-x)} \left[ 5-a +2(2-a)x]+ e_2
  [1+(2-a)x \right]^{3/2},
\end{equation}
where we have introduced $e_1$ and $e_2$ as new arbitrary
constants.

Consequently the solution to the system (\ref{3})-(\ref{5}) can
be written as
\begin{eqnarray}
e^{2\lambda} &=& \frac{1-(a-2)x}{1-x}, \\
e^{2\nu} &=& A^2y^2 , \\
\frac{\rho}{C} &=& \frac{(3-a)[3+(2-a)x]}{[1+(2-a)x]^2}, \\
\frac{p_r}{C} &=& 4 \frac{(1-x)}{[1-(a-2)x]}\frac{\dot{y}}{y}-\frac{3-a}{[1+(2-a)x]}, \\
p_t &=& p_r+\Delta ,\\
\Delta &=& \frac{a(a-2)Cx}{[1-(a-2)x]^2},
\end{eqnarray}
where $y$ is given in (\ref{25}).

Again, if we set $a=0, ~C=1/R^2$ and  $\displaystyle \tilde{x}=\sqrt{1-x}=\sqrt{1-\frac{r^2}{R^2}}$, the solution (\ref{25}) reduces to
\begin{equation}
\label{26} y= e_3 \tilde{x} \left(1- \frac{4}{9}
\tilde{x}^2\right) +e_4 \left(1- \frac{2}{3}\tilde{x}^2\right)^{3/2},
\end{equation}
which is the Tikekar model \cite{Vaidya83} for a superdense neutron star for a different value  of the spheroidal
parameter $K=-2$ (compare with Case I above).

\subsection{Case III: $a-K=-7/4$,~$\alpha =-1/4$}
In this case  (\ref{19}) becomes
\begin{equation}
\label{27} Y =C_1 F\left(-\frac{1}{4}, -\frac{3}{4}, -\frac{1}{2};X\right) + C_2X^{3/2}F\left(\frac{5}{4}, \frac{3}{4},
\frac{5}{2}; X\right),
\end{equation}
which takes the form
\begin{equation}
\label{28} Y = C_1 X^{\frac{3}{2}} \left[1+\sqrt{1-X}\right]^{-3/2} + C_2 \left[1+\sqrt{1-X}\right]^{3/2},
\end{equation}
 and subsequently the general solution is obtained as
\begin{equation}
\label{29} y = h_1 [4-(7+4a)x]^{3/2} \times\left[\sqrt{3+4a}+\sqrt{(7+4a)(1-x)]}\right]^{-3/2} + h_2\left[\sqrt{3+4a}+\sqrt{(7+4a)(1-x)]}\right]^{3/2},
\end{equation}
where $h_1$ and $h_2$ are new arbitrary constants.

Consequently the solution to the system (\ref{3})-(\ref{5}) becomes
\begin{eqnarray}
e^{2\lambda} &=& \frac{4-(4a+7)x}{4(1-x)},\\
e^{2\nu} &=& A^2y^2, \\
\frac{\rho}{C} &=&\frac{(3+4a)[(7+4a)x-12]}{[4-(7+4a)x]^2},\\
\frac{p_r}{C} &=& 16 \frac{(1-x)}{[4-(4a+7)x]}\frac{\dot{y}}{y}+\frac{3+4a}{[4-(4a+7)x]},\\
p_t &=& p_r+\Delta ,\\
\Delta &=& \frac{4a(4a+7)Cx}{[4-(4a+7)x]^2},
\end{eqnarray}
where $y$ is given in equation (\ref{29}).

Even though this new solution does not seem to have any known isotropic analogue, its isotropic limit can be obtained by setting $a=0$.

\section{\label{sec6} Physical analysis}
It is noteworthy that our approach provides a general technique to extend the Tikekar stellar model to the case of anisotropic stellar bodies, and thereby provides a simple mechanism to analyze the impact of anisotropy on stellar properties. To illustrate how the solutions can be used to the investigate the physical behaviour of anisotropic stellar objects, let us consider a particular class of solutions. For a direct comparison, we make use of the transformation equations  
$$C=1/R^2,~~\displaystyle \tilde{x} = \sqrt{1-x} = \sqrt{1-\frac{r^2}{R^2}},$$
and consider the Case II which is an anisotropic extension of the Tikekar superdense star for $K=-2$. Note that the solution has  four unknowns namely, $a$, $R$, $e_3$ and $e_4$. The constant $a$ fixes the extent of anisotropy. The remaining constants can be determined by making use of the boundary conditions (matching of the interior solution to the Schwarzschild exterior solution and demanding that radial pressure should vanish as some finite distance $b$) given below:
\begin{eqnarray}
e^{2\nu} &=& \left(1-\frac{2M}{b}\right),\\
e^{2\lambda} &=& \left(1-\frac{2M}{b}\right)^{-1},\\
p_r (r=b) &=& 0.
\end{eqnarray}

\begin{figure}
\centering
\resizebox{0.75\textwidth}{!}{\includegraphics{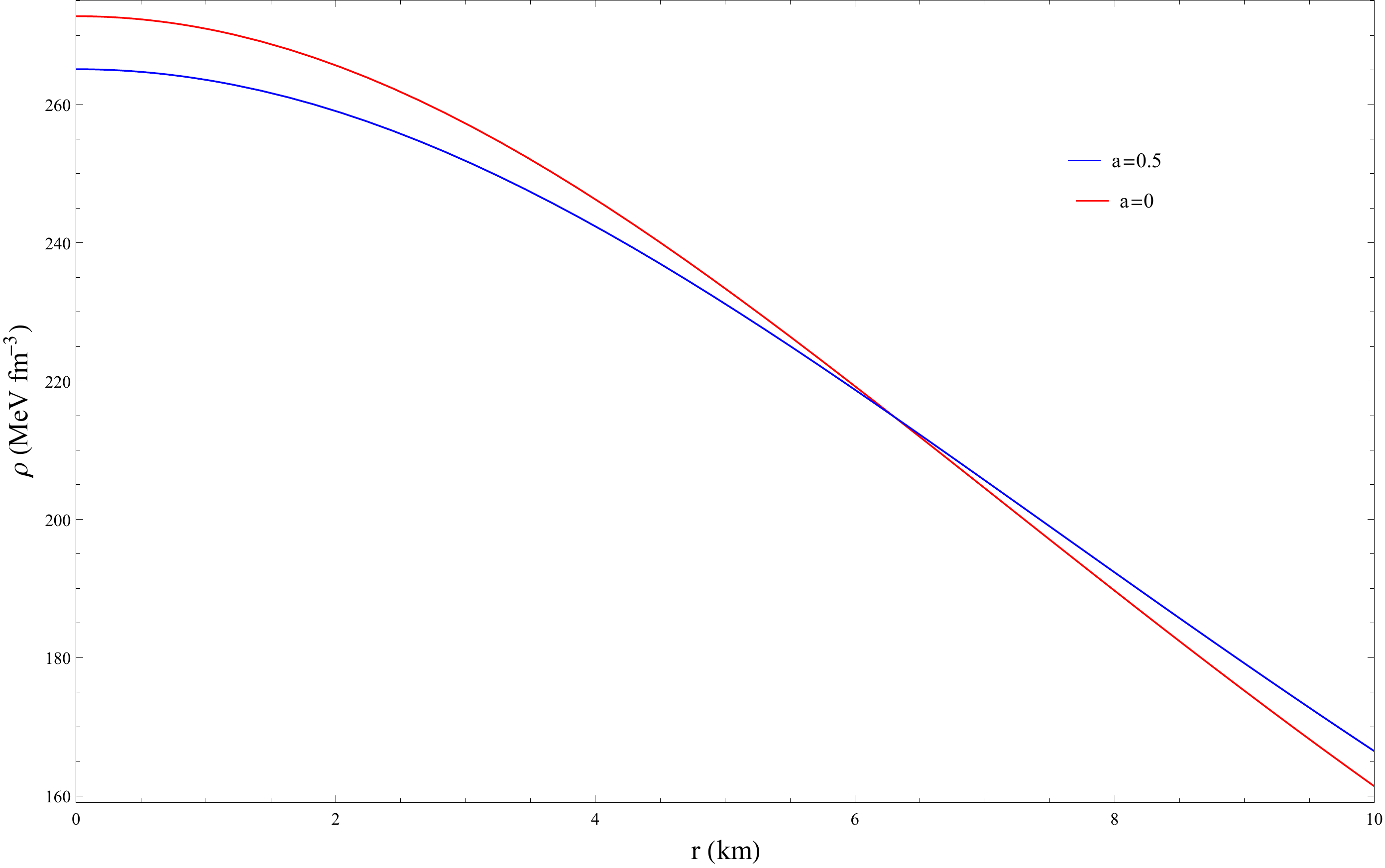}
}
\caption{Density fall-off behaviour.}
\label{fig1}
\end{figure}

\begin{figure}
\centering
\resizebox{0.75\textwidth}{!}{\includegraphics{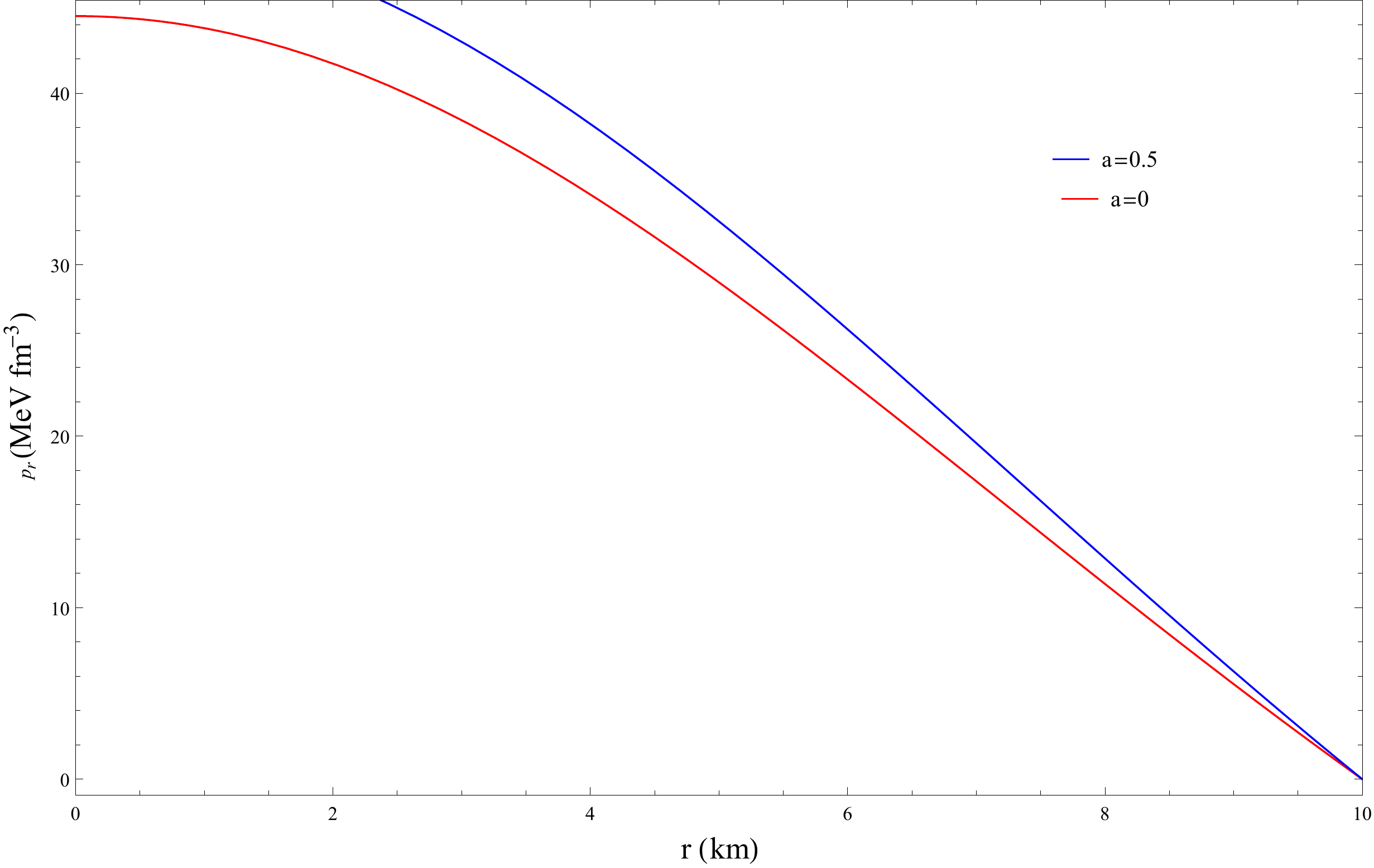}
}
\caption{Radial pressure fall-off behaviour.}
\label{fig2}
\end{figure}

\begin{figure}
\centering
\resizebox{0.75\textwidth}{!}{\includegraphics{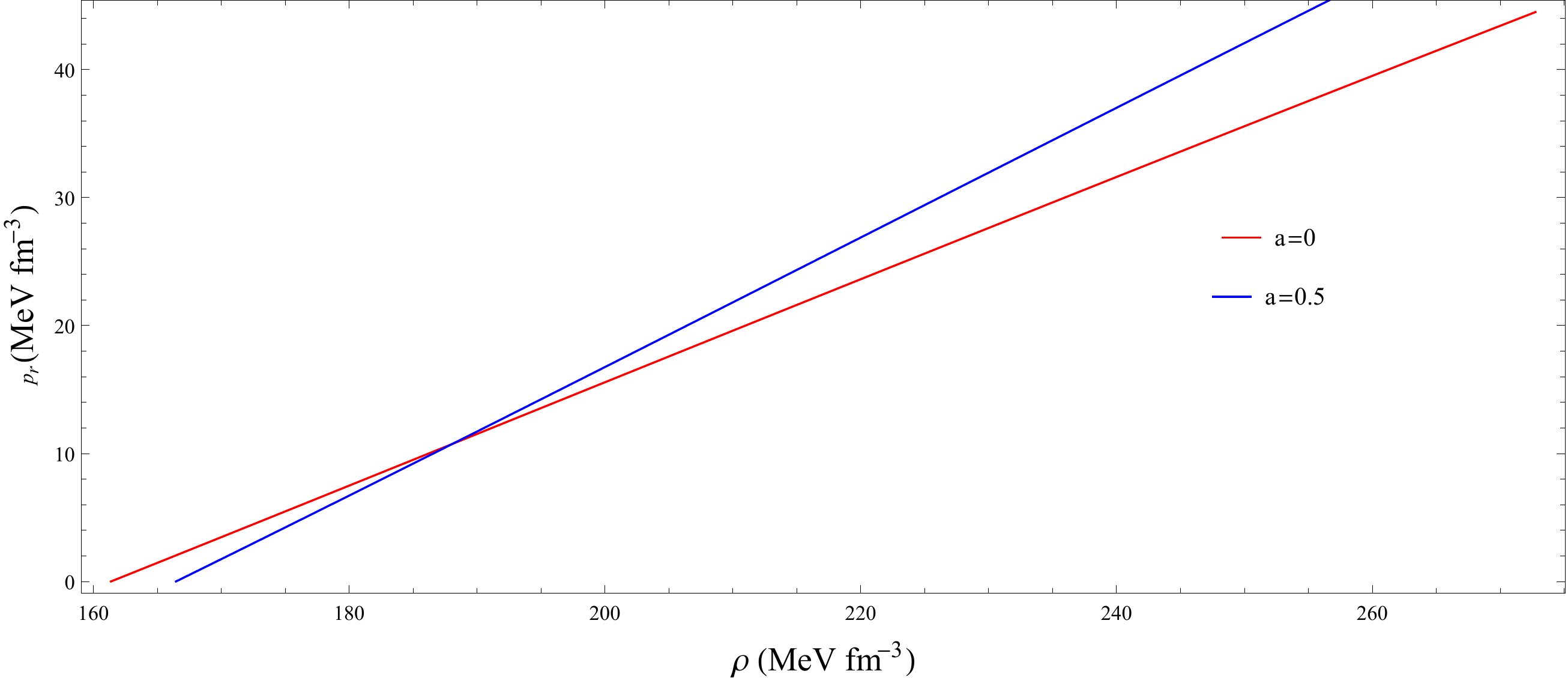}
}
\caption{Density-radial pressure variation (EOS).}
\label{fig3}
\end{figure}

We assume a typical compact star of mass $M= 1.4~M_{\odot}$ and radius $b= 10~$km. Assuming the fluid distribution of the star to be isotropic ($a=0$), the constants are calculated as $R=31.63~$km, $e_3=0.088$ and $e_4=0.1913$. If the composition is further assumed to have anisotropic stress (for which we have assumed $a=0.5$), the constants are obtained as $R=29.28~$km, $e_3=0.0856$ and $e_4=0.2454$. Note that for numerical calculations, we have re-introduced $8\pi G$ and $c$ at the appropriate places. In Fig.~\ref{fig1}, we note that at the central region, the density of an anisotropic star is less than its isotropic counterpart. However, it is just the opposite as one approaches the surface region. The radial pressure of the anisotropic star remains greater than the isotropic counterpart throughout the star as can be seen in Fig.~\ref{fig2}. In Fig.~\ref{fig3},  we have  plotted the density-radial pressure variation inside the star which shows that the EOS becomes comparatively stiffer in the presence of anisotropy.   

\section{\label{sec7} Discussion}
Exact solutions to Einstein field equations concerning various astrophysical and cosmological systems and physical interpretation of the solutions are of immense significance in the theories of gravity\cite{Krasinsky}. In this paper, we have successfully developed a method to generalize a very popular and well studied stellar model by incorporating anisotropic stress into the system. By considering particular spheroidal parameters we can regain known isotropic models e.g., we find the Tikekar superdense stellar models with vanishing anisotropy. Other anisotropic models are also possible. The advantage of our approach is that the generated solutions are not necessarily confined to anisotropic extensions of `known' isotropic solutions. It should be stressed here that the description of a realistic star demands a detailed understanding of the particle interactions at the interior of a compact star. Prescription of the EOS for various layers of quark, hardonic, mixed or some other exotic phases that might exist at the interior of a compact star plays a crucial role in developing a stellar model. Unfortunately, we are still in the process of constraining the EOS in the regime of ultra-high density. Another important aspect that is lacking in our formulation is the spin of a compact star. However, analytical modelling of a non-spherical stellar configuration has its own limitations due to issues relating to matching conditions. Nevertheless, our geometric approach provides a simple way to study the impact of anisotropy on the gross physical properties of relativistic compact stars.  It is noteworthy that even though in this paper we have considered a compact star of mass $1.4~M_{\odot}$ for numerical analysis, the model can accommodate a wide range of masses so far as current estimates of masses and radii of compact stars are concerned. To conclude, it is well known  that the geometric part of the field equations ($G_{\mu\nu}$) essentially depends on the right hand side ($T_{\mu\nu}$) of the field equations. However, our approach provides an alternative technique to get some insight into the physical features as well as composition of a compact star if one assumes observables like mass and radius as input parameters.

\section{Acknowledgements}
\noindent We are thankful to the anonymous referee for his constructive suggestions. The work of RS is supported by the MRP grant F.PSW-195/15-16 (ERO) of the UGC, Govt. of India. RS also gratefully acknowledges support from the Inter-University Centre for Astronomy and Astrophysics (IUCAA), Pune, India, under its Visiting Research Associateship Programme.  SDM acknowledges that this work is based on research supported by the South African Research Chair Initiative of the Department of Science and Technology and the National Research Foundation.

\end{document}